# Odd-even phonon transport effects in strained carbon atomic chains bridging graphene nanoribbon electrodes

Hu Sung Kim, Tae Hyung Kim, and Yong-Hoon Kim*

School of Electrical Engineering and Graduate School of Energy, Environment, Water, and Sustainability, Korea Advanced Institute of Science and Technology, 291 Daehak-ro, Yuseong-gu, Daejeon 305-701, Korea.

Based on first-principles approaches, we study the ballistic phonon transport properties of finite monatomic carbon chains stretched between graphene nanoribbons, an $sp$-$sp^2$ hybrid carbon nanostructure that has recently seen significant experimental advances in its synthesis. We find that the lattice thermal conductance anomalously increases with tensile strain for the even-numbered carbon chains that adopt the alternating bond-length polyyne configuration. On the other hand, in the odd-numbered carbon chain cases, which assume the equal bond-length cumulene configuration, phonon conductance decreases with increasing strain. We show that the strong odd-even phonon transport effects originate from the characteristic longitudinal acoustic phonon modes of carbon wires and their unique strain-induced redshifts with respect to graphene nanoribbon phonon modes. The novel phonon transport properties and their atomistic mechanisms revealed in this work will provide valuable guidelines in designing hybrid carbon nanostructures for next-generation electronic, bio, and energy device applications.

# 1. Introduction

Representing the ideal one-dimensional (1D) $sp^1$-hybridized carbon systems, monoatomic carbon chains (CCs) were predicted to exhibit intriguing physical and chemical properties but their experimental investigations have been relatively slow compared to other $sp^2$- and $sp^3$-carbon allotropes.[1-4] While the infinite CC or carbyne is not yet observed and its existence still remains controversial, much progress has been recently made in the experimental realization of finite CC systems. An interesting aspect of this development is that, rather than the direct synthesis of isolated molecular CCs,[5, 6] they were successfully prepared in the $sp^1$-$sp^2$ hybrid carbon structures. A notable case is the CC confined within carbon nanotubes,[7, 8] and the other is the CC stretched between graphene.[9-17]

In particular, the graphene-CC-graphene system assumes an ideal electrode-channel-electrode junction configuration,[9-17] providing unique opportunities to study the quantum transport properties of all-carbon nanodevices. The infinite CCs can adopt cumulene (··· C=C=C=C ···) or polyyne (··· C≡C–C≡C ···) configuration (Fig. 1a) and exhibit metallic or semiconducting properties, respectively.[3, 4] In a notable example, very recently, strain-induced metal-to-insulator (i.e. cumulene-to-polyyne) transition was observed,[14] confirming another recent theoretical prediction.[18] Another promising example theoretically suggested is spintronic applications such as spin filter and spin valve, which will be operated based on the spin-polarized nature of graphene zigzag edge states and can be modulated by the number of carbon atoms within CCs (cumulene or polyyne).[19-21]

In this work, applying an atomistic Green's function method (AGF) based on density functional theory (DFT) calculations, we investigate the strain-dependent ballistic phonon transport properties of CCs bridging graphene nanoribbon (GNR) electrodes (Fig. 1b). Although several theoretical reports on the charge and spin transport properties of GNR-CC-GNR junctions have previously appeared,[19, 22] study on their phonon transport properties is non-existent.[23] More generally, while ballistic electron and spin transports in nanoscale junctions have been extensively studied in the past decade or so, still much less is known about their ballistic phononic heat transport.[24, 25] We here adopt the microscopic AGF theory to describe the thermal resistance across the dimensionally mismatched $sp^1$-$sp^2$ interfaces.[26-28] Moreover, in spite of the high computational cost, we will utilize first-principles atomics forces obtained through DFT because it was shown in a previous study that classical force fields significantly overestimate phonon transmissions across GNR-atomic carbon contacts.[23]

In studying the ballistic phonon transport properties of various GNR-CC-GNR junction models, we particularly focus on the effects of tensile strain as well as the number of carbon atoms within CCs. Strain was predicted to be an important variable that affects the structural and electronic properties of infinite[29-32] as well as finite CCs.[18, 33, 34] Regarding the number of C atoms within finite CCs, while strong odd-even effects were predicted for the structural and electron transport properties of GNR-CC-GNR junctions,[19, 22, 33, 34] it remains to be seen whether an oscillatory behavior also appears in phonon transport and if does what its nature is. We will show that a strong *strain-dependent* odd-even phonon transport effect indeed arises because phonon conductances in the even-numbered (odd-numbered) CC junctions increase (decrease) with tensile strain. The effect is found in both armchair and zigzag GNRs, indicating the robust nature of the effect. The microscopic mechanisms rationalized by the polyynic atomic structure of even-numbered CCs and the strong redshifting behavior of their longitudinal optical (LO) modes with respect to GNR phonon bands will become useful guidelines in designing hybrid carbon nanostructures for advanced device applications.

# 2. Computational method

## 2.1 *Density-functional theory phonon calculations*

Following our earlier works on stretched molecular junctions,[35, 36] we carried out strain-dependent geometry optimizations within the local density approximation (LDA) of DFT implemented in the SIESTA package.[37] Dynamical matrices were obtained with the DFT forces and the small displacement method using the Phonopy code.[38] Norm-conserving pseudopotentials and double-zeta-plus-polarization quality atomic orbital basis sets were adopted. The convergence criterion for atomic forces was set to $10^{-3}$ eV/Å.

## 2.2 *Quantum phonon transport calculations*

For the computation of ballistic phonon transport properties, we used an in-house code that implements the atomistic matrix Green's function (MGF) formalism[23, 26-28] and was developed based on our electronic MGF code.[35, 39, 40] We are concerned with the linear response limit, or when the difference of the electrode 1/2 temperature $T_{1/2}$ is very small, $T_1 - T_2 \ll T \equiv (T_1 + T_2)/2$. Then, after computing the phonon transmission function,

$$T_{\mathrm{ph}}(\omega) = \mathrm{Tr}[\Gamma_1(\omega)G(\omega)\Gamma_2(\omega)G^+(\omega)], \quad (1)$$

where $G$ is the retarded Green's function matrix of the channel region and $\Gamma_{1/2}$ is the broadening matrix resulting from the coupling of the channel with the electrode 1/2, we calculated the lattice thermal conductance according to

$$K_{\mathrm{ph}}(T) = \int_0^\infty \frac{d\omega}{2\pi} \hbar\omega T(\omega) \frac{\partial n}{\partial T}, \quad (2)$$

where $n = [\exp(\hbar\omega/k_B T) - 1]^{-1}$ is the Bose-Einstein distribution function.

## 3. Results and Discussion

**3.1** *Phononic heat transport in infinite CCs.*

We first discuss the strain-dependent variations in the lattice thermal transport properties of infinite CCs (carbyne) and GNRs, which become the basis of analyzing the phonon transport in GNR-CC-GNR junctions. In terms of the atomic structures of infinite carbynes, we obtained within LDA 1.269 Å (1.301 Å) as the C≡C (C–C) bond length in the polyyne form and 1.285 Å as the C=C bond length in the cumulene counterpart (Fig. 1a). Due to Peierls distortion that drives the bond-length alternation and associated opening of a bandgap,[29, 30, 32] the polyyne structure should be energetically more stable than the cumulene counterpart and within our calculations the energy difference between the two was of 1.12 meV per carbon atom. As the GNR electrodes, we adopted the hydrogen-passivated four zigzag-chain zigzag GNR (4zGNR) and seven dimer-line armchair GNR (7aGNR) with the optimized unit cell lattice constants of 2.460 Å and 4.260 Å, respectively (Fig. 1b).[41]

Upon applying tensile strains, we find for the polyene case that the long C–C bond length increases faster than the short C–C bond length, or the preference for the polyyne configuration is enhanced with strain (Fig.1c).[30, 32] Quantifying the degree of polyynicity using the bond-length alternation (BLA) value, defined as the difference between short C≡C and long C–C bond lengths, we find that BLA increases from of 0.032 Å to 0.130 Å and to 0.207 Å as the tensile strain increases from 0 % to 8 % and to 12 % (Fig. 1d). The Peierls instability manifested in BLA is underestimated within LDA, which incorrectly overestimates the tendency of electron delocalization due to its self-interaction error, and the deficiency could be corrected by employing more elaborate hybrid DFT functionals. However, such calculations are prohibitive for the large GNR-CC-GNR junction models, and more importantly one can expect that the qualitative conclusions based on LDA or generalized gradient approximation will not be modified by employing hybrid functionals.[18, 30] Considering the increasing Peierls distortion and strain effects within hybrid functionals, we expect that the strain-dependent odd-even effects for the GNR-CC-GNR junctions reported below will become more prominent in experimental situations.

In Fig. 2, we summarize the characteristics of strain-dependent phonon properties of infinite CCs as well as 4zGNR. Analyzing the spatial decay of dynamical matrix elements (Fig. 2a upper panels), the force interaction range in CCs is found to be much longer than that in 4zGNRs. Particularly, we observe that the interaction is extremely long-ranged for the cumulene chain, obtaining over the $10^{-2}$ eVÅ$^{-2}$amu$^{-1}$ level values for the off-diagonal dynamical matrix elements even at the 150 Å distance compared with the drop to the $10^{-4}$ eVÅ$^{-2}$amu$^{-1}$ level at about 70 Å (15 Å) for the polyyne (4zGNR) counterpart. These characteristics show up in the phonon band structures of the cumulene chain as a Kohn anomaly (over bending of the longitudinal acoustic (LA) phonon dispersion and the phonon softening) at the Γ point of the polyyne Brillouin zone (or the zone boundary at the cumulene Brillouin zone, Fig. 2b). On the other hand, in the polyyne case, the Peierls distortion opens a gap between the LA and LO branches at the zone boundary X point. (Fig. 2a).[3] The frequency ranges of infinite polyyne and cumulene phonon modes are extended up to about 2250 cm$^{-1}$ (Fig. 2a middle panels, dotted lines) and 2240 cm$^{-1}$ (Fig. 2b middle panels, dotted lines), respectively, within our calculations, and these high-frequency phonon modes are the sources of experimentally observed Raman signals in the 1800 – 2300 cm$^{-1}$ spectral region.[42, 43] Note that no other carbon nanostructure have Raman peaks in this region, so they become the characteristic feature of *sp*-hybridized carbons.[3] In terms of thermal transmissions and conductance, we find that the Kohn anomaly-related features result in $T_{\mathrm{ph}}(\omega)$ of cumulene larger than that of polyyne over some frequency ranges (particularly up to ~ 1200 cm$^{-1}$; Figs. 2a and 2b middle right panels, dotted lines) and thus the $K_{\mathrm{ph}}(T)$ values of cumulene enhanced over those of polyyne (Fig. 2a and 2b lower panels, dotted lines).

Next, applying tensile strains on the polyyne chain, we find that the short triple bond length relatively remains constant while the long single bond length linearly increases (Fig. 1c). The large strain-induced decrease in force constants of alternating single bonds results in significant redshifts of LA and LO modes and corresponding reduction of high-frequency range $T_{\mathrm{ph}}(\omega)$ (Fig. 2a middle panels, dot-dahsed and solid lines), resulting in the decrease of $K_{\mathrm{ph}}(T)$ by ~ 38 % with 12 % strain at 300 K (Fig. 2a lower panels, dot-dahsed and solid lines). For the cumulene case, we again observe a large-scale redshift of $T_{\mathrm{ph}}(\omega)$ corresponding to the LA mode in the high-frequency region, and it in turn results in a ~ 27 % reduction of $K_{ph}(T)$ with 12 % strain at 300 K (Fig. 2b, dot-dahsed and solid lines).

For the cases of 4zGNR (Fig. 2c) and 7aGNR (Fig. S1), the frequency ranges of their phonons reaching up to ~ 1680 cm$^{-1}$ and ~ 1690 cm$^{-1}$, respectively, are lower than those of CCs extending up to ~ 2250 cm$^{-1}$. Moreover, due to the structural rigidity of the hexagonal arrangement of $sp^2$ carbon atoms, the degree of strain-induced phonon band flattening or redshift of high-frequency $T_{ph}(\omega)$ is smaller for GNRs, amounting to the $K_{ph}(T)$ reductions of ~ 14% and ~ 11% in the 4zGNR and 7aGNR, respectively.

**3.2** *Phononic heat transport in infinite CCs.*

We now move on to consider the phonon transport properties of CCs stretched between two GNR electrodes, which is the main focus of this work. Different from the infinite carbyne limit, the ideal case where Peierls distortion could be rigorously defined, the structural and electronic properties of finite CCs such as BLA, electronic band gap, and vibrational properties are strongly affected by the length and termination of CCs.[3, 33] Inserting several odd- and even-numbered CCs between 4zGNR or 7aGNR electrodes, we first prepared a series of GNR-CC-GNR junction models (Fig. 1b). Each junction model was once more optimized with DFT and then its dynamical matrices were calculated to extract coherent phonon transport properties. For the strained junction models, the electrode-region GNRs were stretched by $\Delta L$ = 0.8 Å and once again the DFT-phonon MGF calculations were repeated. More computational details can be found in the Computational method section and Fig. S2.

The results summarized in Fig. 3 strikingly show that we obtain the opposite strain-induced $K_{ph}(T)$ variation trends for the even-numbered and odd-numbered CCs. The strong odd-even effect becomes a robust feature at $T > $ ~ 100 K and is observed for varying number of carbon atoms ($N_C$) as well as for both 4zGNR and 7aGNR electrode cases, indicating its robustness and universality in the GNR-CC-GNR junction configuration. The overall reduction of $K_{ph}(T)$ in junction models compared with that in infinite carbynes can be understood in view of the introduction of two GNR-CC $sp^2$-$sp^1$ contacts. In addition, considering the reduction of $K_{ph}(T)$ with increasing tensile strain in carbynes and GNRs (Fig. 2), the strain-induced $K_{ph}(T)$ reduction in odd-numbered CC is reasonable. On the other hand, the strain-induced enhancement of $K_{ph}(T)$ in even-numbered CC cases cannot be understood using the infinite carbyne data, implying it originates from the unique interactions between finite-CC and GNR phonons.

To explain the microscopic mechanisms of the strain-induced increase (decrease) in $K_{ph}(T)$ for even-numbered (odd-numbered) CC junctions, we analyzed the atomic structures, phonon transmissions, and projected density of states (PDOS) at the unstrained and strained conditions shown in Fig. 4. Note that as shown in Fig. 5 we obtain essentially identical results for the 7aGNR case. To begin with, observing the optimized atomic structures of GNR-CC-GNR junctions in detail (see Figs. S3 and S4), we find that the even-numbered (odd-numbered) CCs stretched between GNRs adopt the alternated polyyne-like (equalized cumulene-like) structures. In addition, compared with the infinite carbyne limits, we observe that the finite CCs in junction models are in effectively strained conditions (Figs. 1c and 1d). Moreover, due to the differences in their structural rigidity, we find that the effective strain induced onto the GNR regions is relatively negligible compare to that of CCs.

Moving on to the $T_{ph}(\omega)$ spectra of eight-carbon (8C) wire bridging 4zGNR electrodes, we immediately notice that $T_{ph}(\omega)$ values at $\omega \approx 1500$ cm$^{-1}$ are significantly enhanced with strain (Fig. 4a, blue down arrows). On the other hand, such $T_{ph}(\omega)$ enhancement at $\omega \approx 1500$ cm$^{-1}$ is absent for the 7C chain case (Fig. 4b), explaining the strong strain-induced odd-even effects in $K_{ph}(T)$ of GNR-CC-GNR junctions. The origins of different strain-induced $T_{ph}(\omega)$ changes in the ~ 1500 cm$^{-1}$ frequency region can be understood by analyzing the phonon PDOS projected onto CCs. As emphasized earlier, the high-frequency LO phonon modes of polyyne chains are discriminated from those of other carbon-based nanostructures,[3] and particularly they are strongly redshifted with increasing tensile strain (Fig. 2). Although the 8C chain within the unstrained junction is in an effectively strained state with reference to the infinite polyyne case (5.8 % in terms of the bond lengths of central carbon atoms), the 8C chain LO mode is still located above the 4zGNR phonon states (Fig. 4a, upper panels). However, upon applying additional strain to the 4zGNR-8C-4zGNR junction, the 8C chain LO mode is further redshifted, making the spectral ranges of the 8C chain and 4zGNR phonons overlap with each other and accordingly increasing $T_{ph}(\omega)$ at $\omega \approx 1500$ cm$^{-1}$ significantly (Fig. 4a, lower panels).

The spatially well-delocalized nature is a necessary condition for a transmission eigenmode to support efficient quantum transport, and the visualization of eigenstates of the molecular projected Hamiltonian has well established such a criterion in the electron quantum transport case.[44, 45] In a similar spirit, we visualized the vibrational modes (phonon local PDOS) corresponding to the enhanced transmission peak at $\omega \approx 1490$ cm$^{-1}$ in the unstrained and strained 8C junction cases, and they indeed clearly show the absence and activation of CC phonon modes in the former and latter cases, respectively (Fig. 4a, right panels). For the 7C chain case, on the other hand, such strain-induced transition behavior is less noticeable (Fig. 4b, right panels). The 7C chain mode at ω ≈1490 cm$^{-1}$ is already activated within the unstrained 4zGNR-8C-4zGNR junction configuration, and correspondingly the transmission and CC PDOS peaks are present in Fig. 4b upper left panels. Such difference between even- and odd-number CC cases can be simplistically understood in terms of the slightly more strained condition for the odd-numbered CCs with reference to the infinite cumulene limit (6.5 % in terms of the bond lengths of central carbon atoms). However, more detailed analysis indicates that odd-numbered CC cases correspond to a complex situation where the mapping to cumulene is not strictly valid (e.g. the C-C bonds adjacent to GNRs are single-bonded rather than double-bonded; see Figs. S3 and S4), and we believe it deserves more close examinations in combination with the electronic structures of different $sp$-$sp^2$ contacts in the future.[17, 33]

## 4. Conclusions

In summary, we carried out *ab initio* phonon transport calculations on finite monatomic CCs stretched between GNR electrodes. Representing the ultimate all-carbon junction model, these hybrid carbon nanostructures based on $sp$-$sp^2$ interfaces are appealing in many aspects but the research efforts have been so far limited to their structural and electronic (and spin) transport properties. Systematically considering a series of CC channels and both zGNR and aGNR electrodes, we found that the phonon conductance of the $sp$-$sp^2$ hybrid structures exhibits strong odd-even oscillations. Specifically, while odd-numbered CCs adopted the cumulenic structure and showed the phonon conductance decrease with increasing tensile strain, even-numbered CCs assumed the polyynic configuration and anomalously showed the strain-induced conductance increase. The strong odd-even ballistic phonon transport effects were rationalized in terms of the strain-dependent behavior of polyyne LO phonon modes, which are characteristic to $sp$-hybridized CCs and exhibit stronger redshifts than the high-frequency phonon modes of $sp^2$ carbons in GNRs. Combined with the electron and spin transport properties that were also predicted to show oscillatory behaviors,[19, 22] the strong odd-even phonon transport effects could open up novel opportunities in advanced nano-device applications.

## Acknowledgements

This work was supported by the Nano-Material Technology Development Program (Nos. 2016M3A7B4024133 and 2016M3A7B4909944), Basic Research Program (No. 2017R1A2B3009872), Global Frontier Program (No. 2013M3A6B1078881), and Basic Research Lab Program (No. 2016M3A7B4909944) of the National Research Foundation funded by the Ministry of Science and ICT of Korea.
## Corresponding Author Information

\*Y.-H.K.: y.h.kim@kaist.ac.kr

## Keywords

phonon quantum transport, carbon atom chains, graphene nanoribbon, first-principles simulation

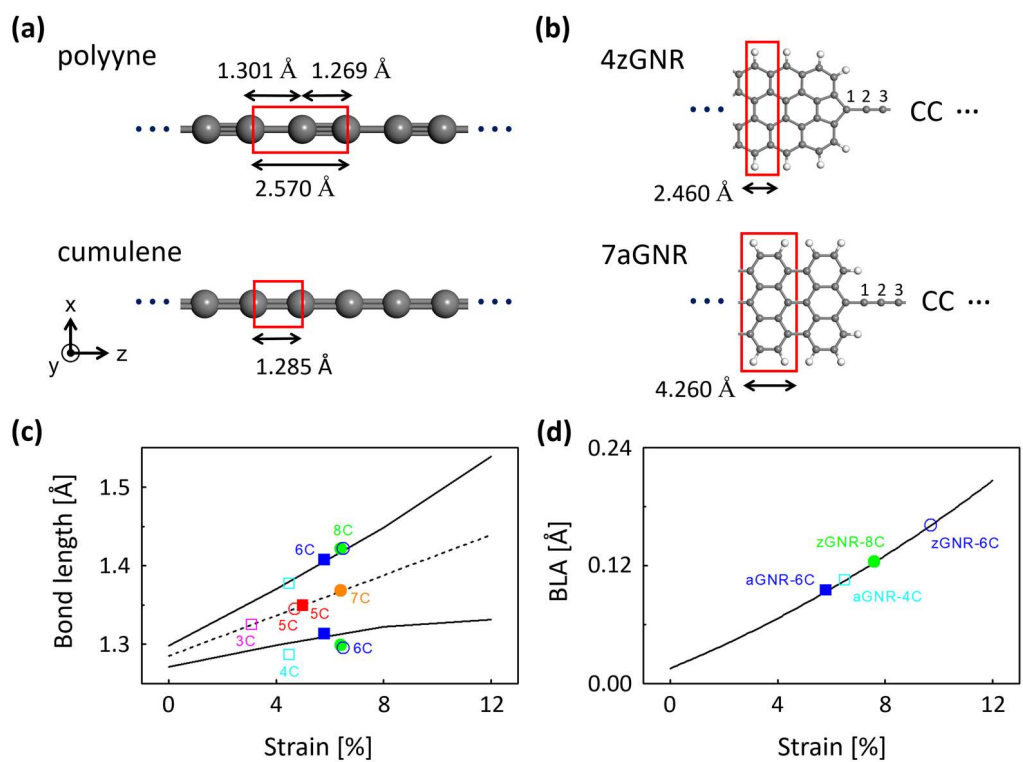

**Figure 1 |** Atomic structures of DFT-optimized (a) polyyne (upper) and cumulene (lower), and (b) 4zGNR and 7aGNR. Schematics of GNR-CC junction contacts are shown together. Here, the red rectangles represent the primitive unit cells. (c) The bond lengths of the finite CC parts within the GNR-CC-GNR junction models are mapped to the bond lengths of infinite polyyne (solid lines) and cumulene (dotted line) at varying strain. Circles (squares) represent 4zGNR (7aGNR) cases. (d) The BLA values of even-numbered CC junctions mapped onto the infinite polyyne BLA values at varying strain. For the junction models, the bond length and BLA values are measured for the innermost carbon atoms.

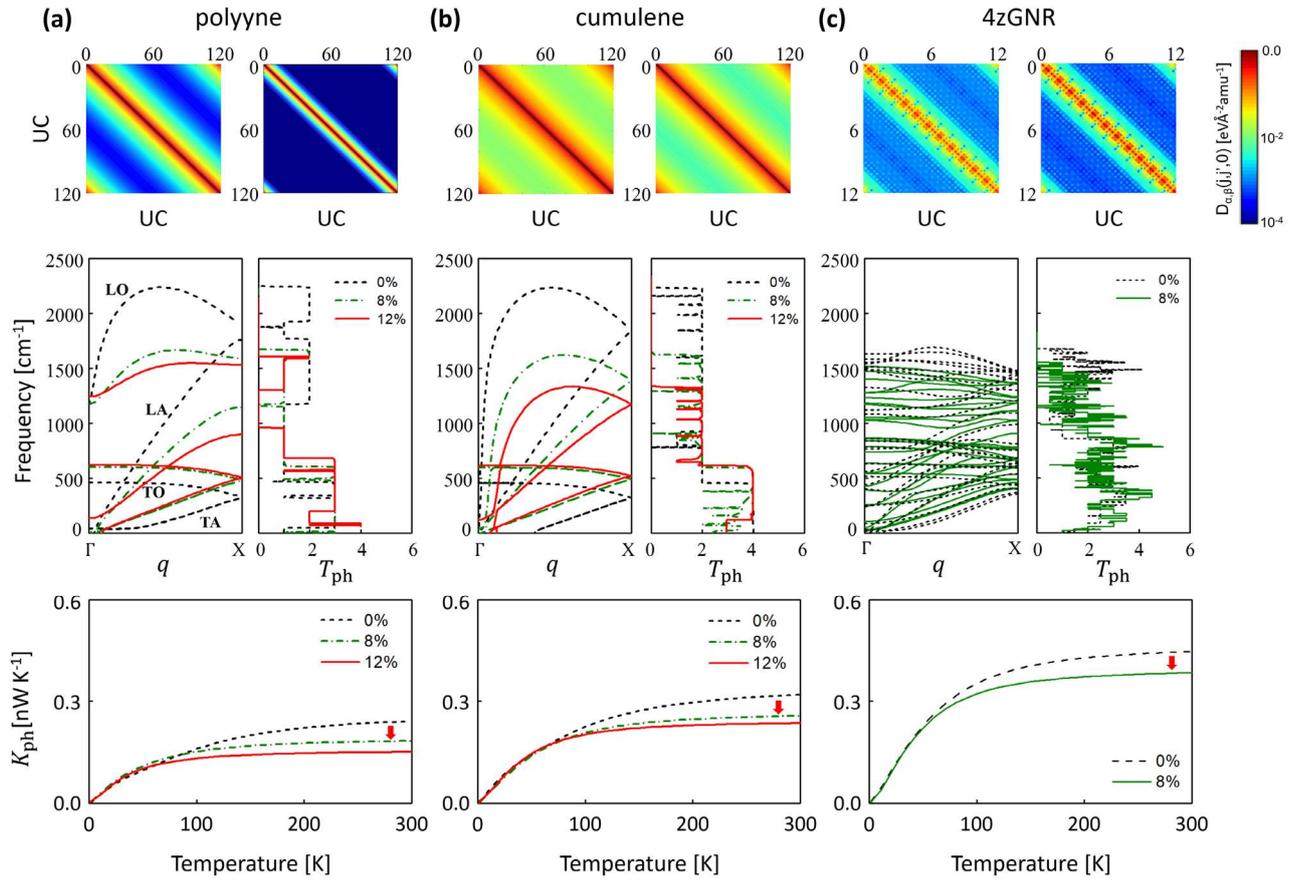

**Figure 2** | Density plots of the dynamical matrices (upper panels), strain-dependent phonon band structures (middle left panels), phonon transmissions (middle right panels), and thermal conductance (lower panels) of (a) polyyne, (b) cumulene, and (c) 4zGNR. For the density plots, left and right panels are from 0% and 8% strain conditions, respectively. Each axis represents the number of unit cells (UCs). The magnitude of matrix elements are given in the units of eVÅ$^{-2}$amu$^{-1}$. For the polyyne and cumulene cases, black dashed, green dash-dot, and red solid lines represent the 0, 8, and 12% strain conditions, respectively. For the zGNR case, black dashed and green solid lines represent the 0 and 8% strain conditions, respectively.

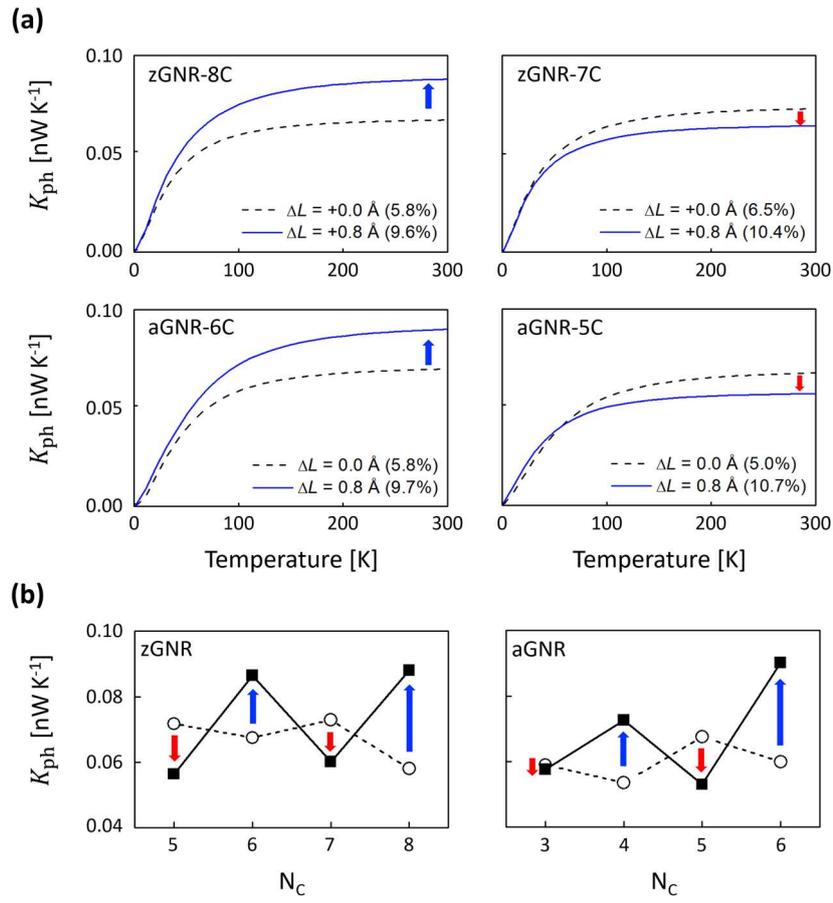

**Figure 3** | Strain-dependent lattice thermal conductances of the representative GNR-CC-GNR junction models. Black dotted and blue solid lines represent the unstrained (dashed lines) and strained (solid lines) conditions, respectively. Effective strain values for the innermost carbon atoms in reference to infinite polyyne and cumulene structures are given for the even-numbered and odd-numbered CC cases, respectively. (b) Compilation of strain-dependent lattice thermal conductance with different number of carbon atoms for the 4zGNR (left) and 7aGNR (right) cases. Black dotted and solid lines represent the unstrained and strained conditions, respectively.

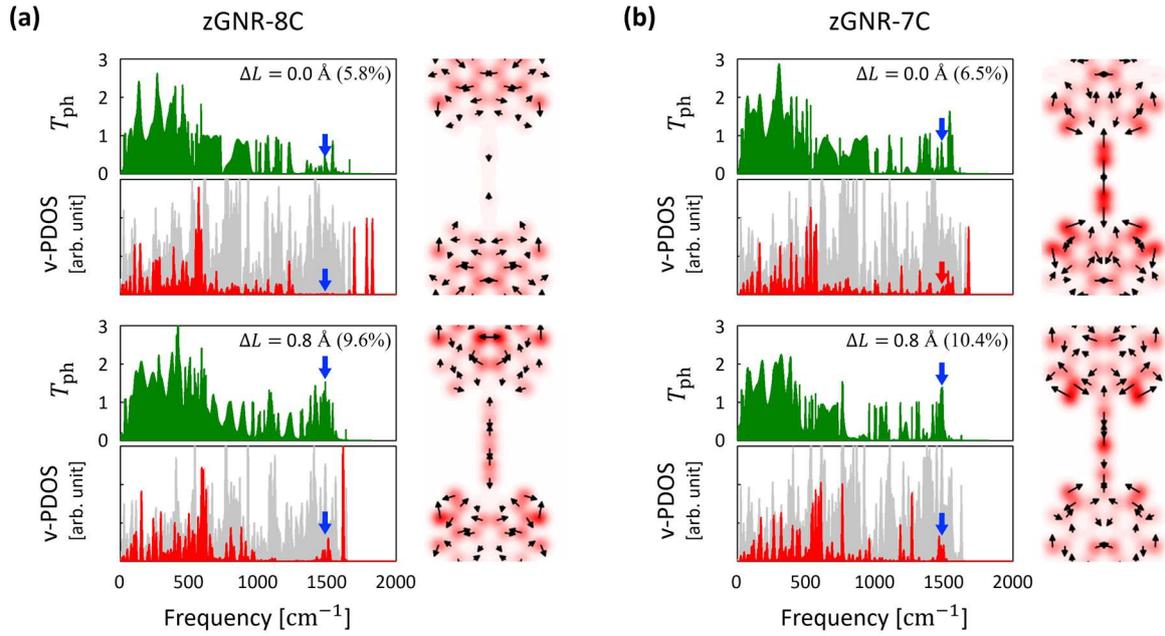

**Figure 4** | Phonon transmissions (left upper panels), vibrational PDOS (left lower panels), and local phonon modes (right panels) at ω ≈ 1490 cm$^{-1}$ in the (a) 4zGNR-8C-4zGNR and (b) 4zGNR-7C-4zGNR junction models for the unstrained (upper panels) and strained (lower panels) conditions. Effective strain values for the innermost CC atoms in reference to infinite polyyne and cumulene structures are given for the 8C and 7C cases, respectively. In the phonon PDOS red and grey lines represent the CC and GNR PDOS, respectively. In the transmission and PDOS data, blue down arrows indicate the frequency where local phonon mode is visualized (ω ≈ 1490 cm$^{-1}$). In the local phonon mode plots, black arrows represent the in-plane vibrational directions of each atom.

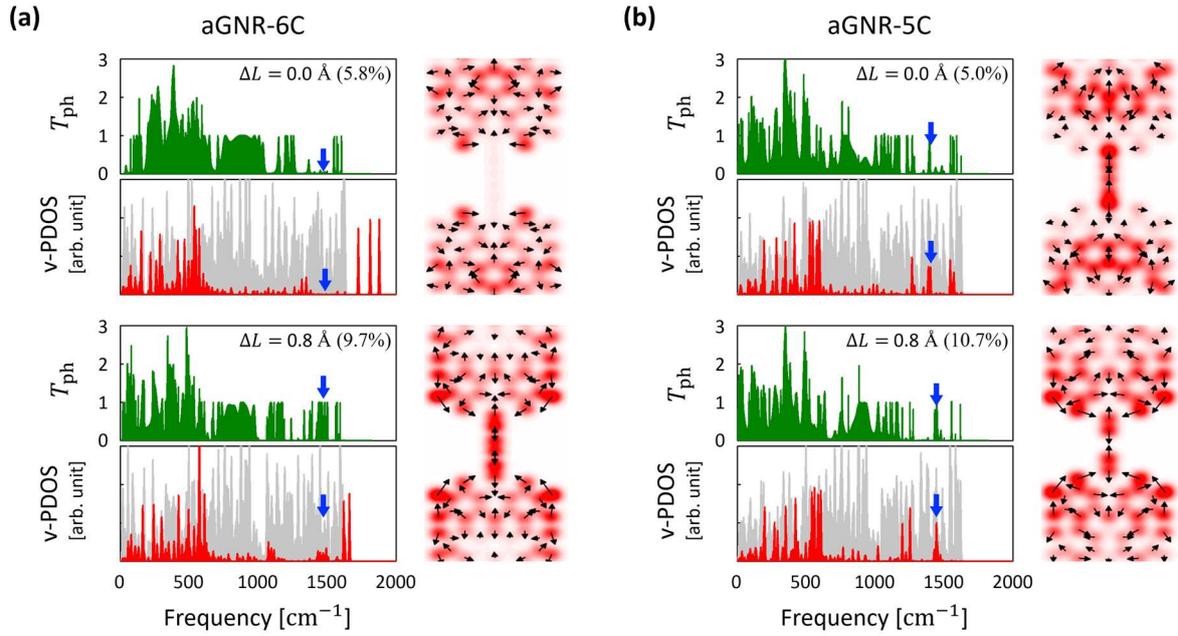

**Figure 5** | Phonon transmissions, vibrational PDOS, and real-space visualizations of ω ≈ 1490 cm$^{-1}$ phonon modes in the (a) 7aGNR-6C-7aGNR and (b) 7aGNR-5C-7aGNR junction models for the unstrained (upper panels) and strained (lower panels) conditions. Effective strain values for the innermost CC atoms in reference to infinite polyyne and cumulene structures are given for the 6C and 5C cases, respectively. Red and grey filled lines in the PDOS plots represent the PDOS of CC and GNR parts, respectively. Blue down arrows indicate the frequency near ω ≈ 1490 cm$^{-1}$. In the local phonon mode plots, black arrows represent the in-plane vibrational directions of each atom.